\documentclass[10pt]{article}
\usepackage{epsfig,graphics,amssymb,amsmath}

\def\s{s$^{-1}$}\def\d{$^\circ$}\def\g{\dot{\gamma}}

\begin{document}

\begin{flushright}
To appear as Ch. 15 in {\it Handbook of Experimental Fluid Dynamics} \\ 
Editors J. Foss, C. Tropea and A. Yarin,  Springer, New-York (2005).

\end{flushright}
\vskip7mm

\begin{center}
{\textbf{ \LARGE 
Microfluidics: The No-Slip Boundary Condition}}
\vskip2mm

{\large  \sc Eric Lauga$^{\dag,}$\footnote{Present address: Department of Mechanical Engineering, Massachusetts 
Institute of Technology, 77 Massachusetts Ave., Cambridge, MA 02139, USA.
 }, Michael P. Brenner$^\ddag$ and Howard A. Stone$^{\dag\dag}$}
\vskip2mm

{\it  Division of Engineering and Applied Sciences, Harvard
University},\\ {\it  29 Oxford Street, Cambridge, MA
 02138,} \\
\vskip2mm
$^\dag$lauga@deas.harvard.edu, 
$^\ddag$brenner@deas.harvard.edu, 
$^{\dag\dag}$has@deas.harvard.edu.
\vskip2mm

September 28, 2005

\vskip6mm

{{\small \bf  Abstract}}
\end{center}

\noindent The no-slip boundary condition at a solid-liquid interface is at the center of our understanding of fluid mechanics. However, this condition is an assumption that cannot be derived from first  principles and could, in theory, be violated. In this chapter, we present a review of recent experimental, numerical and theoretical investigations on the subject. The physical picture that emerges is that of a complex behavior at a liquid/solid interface, involving an interplay of many physico-chemical parameters,  including wetting, shear rate, pressure, surface charge, surface roughness, impurities and dissolved gas.

\vskip8mm

\tableofcontents

\section{Introduction} 

The vast majority of problems in the dynamics of Newtonian fluids are concerned with solving, in particular settings, the Navier-Stokes equations for incompressible flow
\begin{equation}\label{ns}
\rho(\partial_t + {\bf u}\cdot \nabla){\bf u} =- \nabla p + \mu \nabla^2{\bf u},\quad\nabla\cdot{\bf u}=0.
\end{equation}
The list of problems for which this task has proven to be difficult is long. However, most of these studies assume the validity of the no-slip boundary condition, {\it i.e.}, that all three components of the fluid velocity on a solid surface 
 are equal to the respective velocity components of the surface.
  It is only recently that controlled experiments, generally with typical dimensions microns or smaller, have demonstrated an apparent violation of the no-slip boundary condition for the flow of Newtonian liquids near a solid surface.

We present in this chapter a tentative summary of what is known about the breakdown of the no-slip condition for Newtonian liquids and discuss methods and results of experiments, simulations and theoretical models. This topic is of fundamental physical interest and has potential practical consequences in many areas of engineering and applied sciences where liquids interact with small-scale systems \cite{squiresquake,stonereview}, including flow in porous media, microfluidics, friction studies and biological fluids. Furthermore, since viscous flows are relevant to the study of other physical phenomenon, such as the hydrophobic attraction in water, a change in the boundary condition would have significant quantitative impact on the interpretation of experimental results  \cite{vinogradova_short96,vinogradova98,vinogradovahorn01}.

The present chapter complements previous work \cite{granick03,tabeling04,vinogradova99} and is organized as follows. In \S \ref{history} we present a brief  history of the no-slip boundary condition for Newtonian fluids, introduce some terminology, and discuss cases where the phenomenon of slip (more appropriately, this may often be ``apparent slip'') has been observed. In  \S\ref{methods} we present the different experimental methods that have been used to probe slip in Newtonian liquids and summarize their results in the form of tables. A short presentation of the principle and results of Molecular Dynamics simulations is provided in \S\ref{MD}, as well as remarks about the relation between simulations and experiments. We then present in \S\ref{discussion} an interpretation of experimental and simulation results in light of both molecular and continuum models, organized according to the parameters upon which slip has been found to depend. We conclude in \S\ref{perspective} by offering a brief perspective on the subject.

\section{History of the no-slip condition} 
\label{history}

\subsection{The previous centuries} 

\begin{figure}[t]
\centering
\includegraphics[width=.7\textwidth]{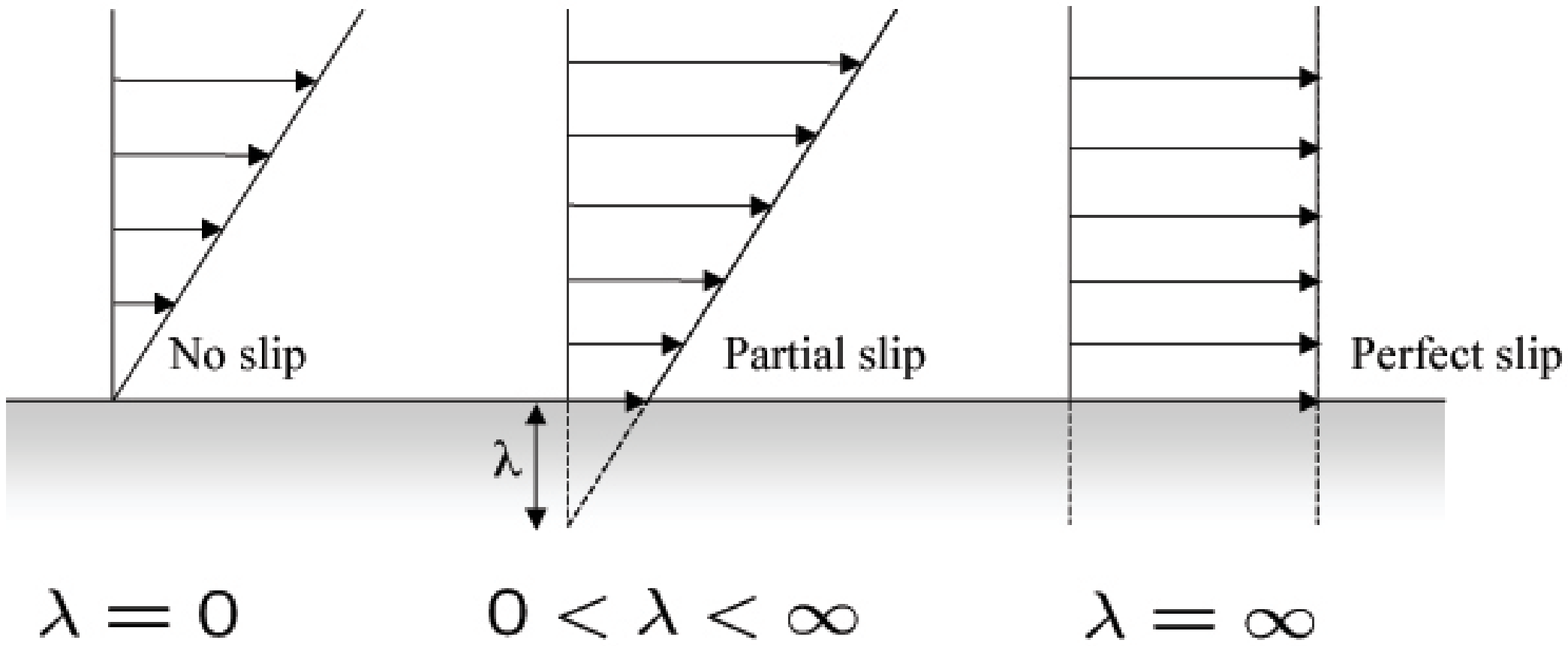}
\caption{\label{navier} Intepretation of the (Maxwell-Navier) slip length $\lambda$.}
\end{figure}

The nature of boundary conditions in hydrodynamics was widely debated in the 19th century and the reader is referred to \cite{goldstein,goldstein69} for historical reviews. Many of the great names in fluid dynamics have expressed an opinion on the subject at some point during their careers, including D. Bernoulli, Euler, Coulomb, Darcy, Navier, Helmoltz, Poisson, Poiseuille, Stokes, Hagen, Couette, Maxwell,  Prandtl and Taylor. In his 1823 treatise on the movement of fluids \cite{navier}, Navier introduced the linear boundary condition (also proposed later by Maxwell \cite{maxwell79}), which remains the standard characterization of slip used today: the component of the fluid velocity tangent to the surface, ${\bf u}_\parallel$, is proportional to the rate of strain, (or shear rate) at the surface,\footnote{Alternatively, the right-hand side of this boundary condition, when multiplied by the shear viscosity of the liquid, states that the tangential component of the surface velocity is proportional to the surface shear stress.}
\begin{equation}\label{bc}
 {\bf u}_\parallel =\lambda {\bf n}\cdot  ({\nabla}{\bf u}+({\nabla}{\bf u})^T)\cdot ({\bf 1}-{\bf n}{\bf n}),
\end{equation}
where  ${\bf n}$ denotes the normal to the surface, directed into the liquid. The velocity component normal to the surface is naturally zero as mass cannot penetrate an impermeable solid surface,  ${\bf u}\cdot{\bf n} =0$. In Eq.~\eqref{bc}, $\lambda$ has the unit of a length, and is referred to as the slip length. For a pure shear flow, $\lambda$ can be interpreted as the fictitious distance below the surface where the no-slip boundary condition would be satisfied (see Fig.~\ref{navier}). Note that on a curved surface the rate of strain tensor is different from the normal derivative of the tangential component of the flow so all the terms in Eq.~\eqref{bc} need to be considered \cite{einzel90}.

A century of agreement between experimental results in liquids and theories derived assuming the  no-slip boundary condition ({\it i.e.}, $\lambda = 0$) had the consequence that today many textbooks of fluid dynamics  fail to mention that the no-slip boundary condition remains an  assumption. A few monographs however discuss the topic. In his classic book  \cite{lamb} (p. 576), Lamb realizes that no-slip is the most probable answer but leaves the possibility open for extraordinary cases:
\begin{quote}
{\small It appears probable that in all ordinary cases there is no motion, relative to the solid, of the fluid immediately in contact with it. The contrary supposition would imply an infinitely greater resistance to the sliding of one portion of the fluid past another than to the sliding of the fluid over a solid.}
\end{quote}
Similarly, Batchelor \cite{batchelor} (p. 149) offers two paragraphs where the question is discussed in detail, including the role of molecular effects in smoothing out discontinuities. He also mentions the importance of an experimental validation of the no-slip condition:
\begin{quote}
{\small 
The validity of the no-slip boundary condition at a fluid-solid interface was debated for some years during the last century, there being some doubt about whether molecular interactions at such an interface lead to momentum transfer of the same nature as that at a surface in the interior of a fluid; but the absence of slip at a rigid wall is now amply confirmed by direct observations and by the correctness of its many consequences under normal conditions.}
\end{quote}

\subsection{Terminology}
We introduce here some useful terminology that is used throughout this chapter.

\begin{itemize}

\item {\bf Phenomenon of slip}: Refers to any situation in the dynamics of  fluids where the value of the tangential component of the velocity appears to be different from that of the solid surface immediately in contact with it.

\item {\bf Molecular slip} (also intrinsic slip): Refers to the possibility of using hydrodynamics to force liquid molecules to slip against solid molecules. Such a concept necessarily involves large forces \cite{tabeling04}. Let us denote by $\sigma$ a typical molecular length scale and by  $A$  the Hamaker constant for the intermolecular forces. Molecular slip will occur when intermolecular interactions ${\cal O}( A / \sigma)$ are balanced by viscous forces ${\cal O}( \mu \sigma^2\g)$ where $\mu$ is the shear viscosity of the liquid and $\g$ the shear rate; this  can only happen for a very large shear rates $\g\approx A/\mu \sigma^3 \sim 10^{12}$ \s, where we have taken the viscosity of water $\mu=10^{-3}$ Pa$\cdot$s, and typical values $A\approx 10^{-19}$ J and $\sigma\approx 0.3$ nm.

\item {\bf Apparent slip}: Refers to the case where there is a separation  between a small length scale $a$ where the no-slip condition is valid and a large length scale $L\gg a$ where the no-slip condition appears to not be valid. Well-known examples of such apparent slip include electrokinetics \cite{saville} (in this case $a$ is the the thickness of the double layer) and acoustic streaming \cite{batchelor} (in this case $a$ is the thickness of the oscillatory boundary layer). Similarly, a liquid flowing over a gas layer displays apparent slip (see \S\ref{discussion}).

\item {\bf Effective slip}: Refers to the case where molecular or apparent slip is estimated by averaging an appropriate measurement over the length scale of an experimental apparatus.

\end{itemize}
\subsection{Traditional situations where slip occurs} 

The phenomenon of slip has already been encountered in three different contexts.

\paragraph{Gas flow.}

Gas flow in devices with dimensions that are on the order of the mean free path of the gas molecules shows significant slip \cite{muntz89}. An estimate of the mean free path is given by the ideal gas formula $\ell_m\approx  1/(\sqrt{2}\pi \sigma^2 \rho)$ where $\rho$ is the gas density (here taken as the number of molecules per unit volume);  for air under standard conditions of temperature and pressure,  $\ell_m\approx 100$ nm  and, in general, $\ell_m$ depends strongly on  pressure and temperature. The possibility of gas slip was  first introduced by Maxwell \cite{maxwell79}. He considered the flow of an ideal gas and assumed that a percentage ($1-p$) of the wall collisions were specular whereas a percentage ($p$) were diffusive. Such an assumption allows an exchange of momentum between the gas and the wall. The corresponding slip length is given by
\begin{equation}
\frac{\lambda}{\ell_m}=\frac{2(2-p)}{3p}\cdot
\end{equation}
The case of a rough surface with only specular reflections was considered in \cite{bocquet93}.
In general, a Knudsen number defined as the ratio of the mean free path to the system size
${\rm Kn}=\ell_m/L$ is used to characterize the boundary condition for gas flow, with slip being important when ${\rm Kn}\gtrsim 0.1$ \cite{gadelhak}.

\paragraph{Non-Newtonian fluids.}

The flows of non-Newtonian fluids such as polymer solutions show significant apparent slip in a variety of situations,  some of which can lead to slip-induced instabilities. This is a topic with a long history and is of tremendous practical importance, and  we refer to Refs.~ \cite{brochard92,degennes79,denn01,inn96,kraynik81,leger99,migler93,schowalter88,wang99} and references therein for an appropriate treatment. 

\paragraph{Contact line motion.}

In the context of Newtonian liquids, molecular slip has been used as a way to remove singularities arising in the motion of contact lines, as reviewed in \cite{degennes85,dussan79}. Solving the equations of motion  with a no-slip boundary condition  in the neighborhood of a moving contact line leads to the conclusion that the viscous stresses and the rate of energy dissipation have non-integrable singularities. It was first  suggested \cite{huh71} that a local slip boundary condition for the flow would remove the singularity, and indeed it does \cite{hocking76,dussanv76} (see also \cite{dussan74}). Furthermore, since the slip length appears via a logarithmic factor in a condition involving the contact line speed, it has  virtually no influence on macroscopic quantities such as force and pressure drops on scales larger than  the capillary length $\ell_c=(\gamma/\rho g)^{1/2}$ \cite{dussanv76}, where $\gamma$ is the liquid surface tension, $\rho$ the liquid density and $g$ gravity. 
Consequently, macroscopic measurements on moving droplets cannot be used in general to deduce the exact slip law \cite{eggersstone}. The slip length could also become velocity dependent due to a combination of microscopic roughness and contact angle hysteresis \cite{jansons86}.
Such local slip near contact lines was confirmed by early Molecular Dynamics simulations  \cite{koplik88}. In a different context but with a similar purpose, slip was used to remove the singularity in the mobility of particles sliding near solid surfaces \cite{davis94,hocking73}.

\subsection{Newtonian liquids: No-slip? Slip?} 

The development of the Surface Force Apparatus in the 1970's \cite{israelachvili78,israelachvili72,tabor69} (see \S\ref{SFA}) has allowed for more than thirty years of  precise probing down to the nanometer scale  of both structure and dynamics of many Newtonian liquids against mica \cite{horn85,israelachvili90,georges93,horn89,israelachvili86,israelachvili88,klein95,klein98_1,klein98_2,klein02,raviv01}. Experimental methods have included squeeze and/or shear flow for a variety of polar and non-polar liquids displaying a wide range of wetting conditions and shear rates. With the exception of  the flow of toluene over C$_{60}$ (Fullerene)-coated mica \cite{campbell96}, these studies have confirmed the validity of the no-slip boundary condition and the bulk rheological behavior down to a  few nanometers. At smaller length scales, an increase in viscous resistance has been reported, with qualitative differences between the behavior of water and other non-polar liquids \cite{israelachvili90,granick91,klein95,raviv01}. The conclusions have been confirmed by Molecular Dynamics simulations \cite{muser00} and are consistent with studies of flow in capillaries with diameters of tens of nanometers \cite{anderson72_2,knudstrup95}.

In this context, the large number of recent published experiments reporting some form of (apparent) slip with $\lambda\sim 1$ nm$-1$ $\mu$m in the flow of Newtonian liquids is surprising \cite{baudry01,boehnke99,bonaccurso03,bonaccurso02,cheikh03,cheng02,breuer03,churaev02,cottin-bizonne05,cottin02,craig01,henry04,joseph05,churaev99,leger03,lumma03,neto03,pit99,pit00,sun02,meinhart02,meinhart04,vinogradova03,granick01,granick02_macro,granick02,granick02_langmuir}, and has allowed to re-discover a few early studies reporting some degree of slip \cite{bulkley31,churaev84,debye59,schnell56,traube28}. In part, this chapter is an attempt to describe and interpret these more recent experimental results.


\begin{table}[t]\scriptsize
\begin{center}
\begin{tabular}{llllllll}
\hline \\
								& Surfaces	& Liquids			& Wetting	&Roughness 	& Shear rates  & 	Slip length & L/NL \\ \\
Schnell \cite{schnell56} 				&Glass+DDS 	& Water 			& $-$ 		&$-$ 		& $10^2-10^3$ \s 	& $ 1-10$ $\mu$m  &L\\
Churaev  \cite{churaev84} 		&Quartz+TMS 		& Water 			& $70-90$\d 		&$-$ 		& $1$ \s			& $30$ nm  &NL\\
 								& ''		& Mercury 		& $115-130$\d 		&$-$ 		& 	$10^3-10^4$ \s			& $70$ nm  &NL\\
 								& ''		& CCl$_4$ 		& Complete		& $-$		& $-$				& no-slip&$-$\\
 								& ''		& Benzene		& Complete 		& $-$		& $-$				& no-slip&$-$\\
Kiseleva  \cite{churaev99} 		& Quartz+CTA(+) & CTAB solutions & 70\d& $-$& $10^2-10^3$ \s & $10$ nm &L\\
Cheng  \cite{cheng02} 			&Glass+photoresist & Water 		&$-$ 			&5 \AA\, (pp) &$10^2-10^4$ \s & no-slip&$-$\\
					 			&''  				&Hexane		&$-$ 			&'' &'' & $ 10$ nm  &L\\
					 			&'' 				&Hexadecane 	&$-$ 			&'' &'' & $ 25$ nm &L\\
					 			&'' 				&Decane 		&$-$ 			&'' &'' & $ 15$ nm &L\\
					 			&'' 				&Silicon Oil	&$-$ 			&'' &'' & $ 20$ nm &L\\
Cheikh  \cite{cheikh03}			& Poly(carbonate)+PVP	& SDS solutions	&  $<90$\d			& $-$ & $0-10^5$ \s& $ 20$ nm &L\\					Choi  \cite{breuer03} 			& Silicon 		& Water 	& $\approx 0$\d & $11$ \AA\,(rms) & $10^3-10^5$ \s& $ 0-10$ nm &NL\\ 
								& Silicon+OTS & Water 	& $\gtrsim 90$\d &$3$ \AA\,(rms) &''  & $ 5-35$ nm &NL\\ \\
\hline
\end{tabular}
\end{center}
\caption{\label{table:PD} Summary of slip results for pressure drop versus flow rate experiments. The following symbols are used in this table: 
$-$: unknown parameter;  
DDS: dimethyldichlorosilane; 
TMS: trimethylchlorosilane; 
CTAB / CTA(+): cetyltrimethyl ammonium bromide; 
PVP: polyvinylpyridine; 
OTS: octadecyltrichlorosilane; 
CCl$_4$: tetrachloromethane; 
SDS: sodium dodecyl sulfate; 
pp: peak to peak; 
rms: root mean square; 
L: slip independent of shear rate; 
NL: shear rate dependent.}
\end{table}

\begin{table}[ht]\scriptsize
\begin{center}
\begin{tabular}{llllllll}
\hline \\
& 		  Surfaces	&	Liquids	& 	Wetting	&	Roughness 	&  	Shear rates	 & 	Slip length & L/NL\\ \\
S: Boehnke   \cite{boehnke99} 	&  Silica  & Propanediol & $\approx 0$\d &$-$ & 1 \s& no-slip & $-$\\
							&  '' & Propanediol+Va &'' & $-$& ''& $ 1$ $\mu$m &$-$\\
							&  '' & PDMS &$-$ & $-$& ''& no-slip& $-$\\
							&  Silica+DETMDS & Propanediol& $70-80$\d& $-$& ''& no-slip& $-$\\
							&  '' & Propanediol+Va& ''& $-$& ''& $ 1$ $\mu$m &$-$\\
							&  '' & PDMS& $-$&$-$ & ''& no-slip & $-$\\
FR: Pit  \cite{pit99,pit00,leger03} 	&  Saphire& Hexadecane & Complete &4 \AA\,(rms) &$10^2-10^4$ \s & $ 175$ nm & L\\
							 & Saphire+FDS &'' &65\d &'' &'' & no-slip& $-$\\
							 & Saphire+OTS&'' & 40\d &'' &'' & $ 400$ nm & L\\
							 & Saphire+STA&'' & 25\d &'' &'' & $ 350$ nm &L\\
PIV: Tretheway   \cite{meinhart02,meinhart04} &Glass &Water & $\approx 0$\d& $-$& $10^2$ \s & no-slip& $-$\\
									 & Glass+OTS&''  & 120\d& 2 \AA & '' & $ 0.9$ $\mu$m &$-$\\
PIV: Joseph   \cite{joseph05} &Glass &Water & $\approx 0$\d& 5 \AA (rms)& $10^2$ \s & $50$ nm& $-$\\
									 & Glass+OTS&''  & 95\d& '' & '' & no-slip &$-$\\
									 & Glass+CDOS&''  & 95\d& '' & '' & $50$ nm &$-$\\
SP: Churaev   \cite{churaev02}  &Quartz & KCl solutions & $-$& 2 nm (pp) & $10^5$ \s& no-slip& $-$\\
						  & Quartz+TMS & KCl solutions & $80-90$\d& 25 nm (pp)& ''& $ 5-8$ nm &$-$\\	
FC: Lumma   \cite{lumma03}  	  &Mica &Water &$-$ & 15 nm (pp) &10 \s &  $ 0.5-0.86$ $\mu$m &$-$\\ 
 						 &Glass &Water &$5-10$\d & '' &'' & $ 0.6-1$ $\mu$m & $-$\\ 
 						 &'' &NaCl solutions &'' & '' &'' & $ 0.2-0.6$ $\mu$m & $-$ \\
\\
\hline
\end{tabular}
\end{center}
\caption{\label{table:OTHER} Summary of alternative experimental methods to infer slip. 
The symbols  used in this table are given in Table~\ref{table:PD}, with additional symbols as:
S: sedimentation; 
FR: fluorescence recovery; 
PIV: particle image velocimetry; 
SP: streaming potential; 
FC: fluorescence cross-correlations; 
DETMDS: diethyltetramethyldisilazan; 
FDS: perfluorodecanetrichlorosilane; 
STA: stearic acid (octadecanoic acid); 
CDOS: chlorodimethyloctylsilane;
Va: vacuum; 
PDMS: polydimethylsiloxane; 
KCl: potassium chloride; 
NaCl: sodium chloride.
}
\end{table}


\begin{table}[ht]\scriptsize
\begin{center}
\begin{tabular}{llllllll}
\hline \\
							& 	Surfaces		&	Liquids	& 	Wetting	&	Roughness 	&  	Shear rates	 & 	Slip length & L/NL \\ \\
Chan  \cite{horn85} 		& Mica			&OMCTS	& $-$ &$-$  &$10-10^3$  \s & no-slip &$-$ \\
						& ''			&Tetradecane	& $-$ &$-$  &'' & no-slip &$-$ \\
						& ''			&Hexadecane	& $-$ &$-$  &'' & no-slip &$-$ \\
Israelachvili \cite{israelachvili86} 		& Mica 			& Water 	& $-$ & $-$ & $10-10^4$ \s & no-slip &$-$ \\
						 		& '' 			& Tetradecane 	& $-$ & $-$ & ''  & no-slip &$-$ \\
Horn  \cite{horn89} 		& Silica 			& NaCl solutions	& 45\d & 5 \AA (av) & $10-10^3$ \s & no-slip &$-$ \\
Georges  \cite{georges93} & 6 surfaces \cite{georges93}			& 9 liquids  \cite{georges93} 	& $-$ & 0.2 - 50 nm (pp) & $1-10$ \s & no-slip &$-$ \\
Baudry  \cite{baudry01} 		& Cobalt 			& Glycerol 	&$20-60$\d & 1nm (pp) & $1-10^4$ \s &no-slip  & $-$\\
							& Gold+thiol 		& ''			& $90$\d&''  &''  & $ 40$ nm  & L\\
Cottin-Bizonne  \cite{cottin02}	& Glass  			&Glycerol 	& $<5$\d 	& 1nm (pp) 	& $1-10^4$ \s 	& no-slip  & $-$\\
							& Glass+OTS 		&Glycerol & 95\d	& 	'' 		& '' 			& $ 50-200$ nm  & L\\
							& '' 		&Water 	& 100\d	& 		'' 	& 	'' 		& $ 50-200$ nm  & L\\
Zhu  \cite{granick01}			& Mica+HDA  	& Tetradecane & 12\d 	& $\approx 1$ \AA \,(rms) & $10-10^5$ \s & $ 0-1$ $\mu$m  & NL\\
							& Mica +OTE 		& Tetradecane & 44\d 	& '' &''  & $ 0-1.5$ $\mu$m  & NL\\
							& '' 		& Water 		& 110\d 	& '' & '' & $ 0-2.5$ $\mu$m  & NL\\
Zhu  \cite{granick02}			&Mica+OTS &Water  & $75-105$\d& 6 nm (rms) &$10-10^5$ \s &  no-slip &$-$ \\
							&'' &Tetradecane  & $12-35$\d& 6 nm (rms) & '' &  no-slip & $-$\\
							&Mica+.8 PPO &Water  & $85-110$\d& 3.5 nm (rms) & '' & $ 0-5$ nm  & NL\\
							&'' & Tetradecane  & $21-38$\d& 3.5 nm (rms) &''  & $ 0-5$ nm  & NL\\
							&Mica+.2 PPO &Water  & $90-110$\d& 2 nm (rms) &''  & $ 0-20$ nm  & NL\\
							&'' & Tetradecane  & $-$& 2 nm (rms) & '' & $ 0-20$ nm  & NL\\
							&Mica+OTE & Water & $110$\d&  0.2 nm (rms) &''  & $ 0-40$ nm  & NL\\
							&'' & Tetradecane & $38$\d&  0.2 nm (rms) & '' & $ 0-40$ nm  & NL\\
Zhu  \cite{granick02_macro}	&Mica+PVP/PB 	& Tetradecane & $-$& $\approx 1$ nm (th)&$10-10^5$ \s & no-slip & $-$\\
							&Mica+PVA 		& Water &$-$ & ''&$10-10^5$ \s & $ 0-80$ nm  & NL\\
Zhu  \cite{granick02_langmuir}	&Mica& n-Alkanes& Complete &$-$ &$10-10^5$ \s & no-slip  & $-$\\ 
							&Mica+HDA & Octane & $-$&$-$ &$10-10^5$ \s & $ 0-2$ nm  & NL \\
							&'' & Dodadecane& $-$&$-$ & ''  & $ 0-10$ nm  & NL \\ 
							&'' & Tetradecane & 12\d&$-$ & ''  & $ 0-15$ nm  & NL \\
Cottin-Bizonne  \cite{cottin-bizonne05}	&  	Glass		& Dodecane 	&$\approx$ 0 \d 	& 1 nm (pp) 	& $10^2-10^4$  \s 	& no-slip  &$-$ \\
									&  	''			& Water		& $\approx$ 0\d 	&  ''	&  '' 	& no-slip  &$-$ \\
									&  	Glass+OTS	& Dodecane	& $-$ 			&  ''	& '' 	& no-slip  &$-$ \\
									&  	''			& Water		& 105\d 			&  ''	&  '' 	& 20 nm  & L \\ \\							
\hline
\end{tabular}
\end{center}
\caption{\label{table:SFA} Summary of slip results for experiments using the Surface Force Apparatus (SFA). 
The symbols  used in this table are given in Tables~\ref{table:PD}  and  \ref{table:OTHER}, with additional symbols as:
HDA: 1-hexadecylamine; 
OTE: octadecyltriethoxysilane; 
PPO:  polysytrene (PS) and polyvinylpyridine (PVP), followed by coating of OTE;  
PVP/PB: polyvinylpyridine and polybutadiene; 
PVA: polyvinylalcohol; 
OMCTS: octamethylcyclotetrasiloxane; 
av: average; 
th: polymer thickness. 
Note that many entries in this table, including the largest slip lengths, are from the same group (S. Granick, U. Illinois).}
\end{table}


\begin{table}[ht]\scriptsize
\begin{center}
\begin{tabular}{llllllll}
\hline \\
								&Surfaces	&	Liquids	& 	Wetting	&	Roughness 	&  	Shear rates	 & 	Slip length & L/NL \\ \\
Craig  \cite{craig01} 				& Silica+gold+thiols& Sucrose sol.&$40-70$\d &6 \AA\, (rms) & $10-10^6$ \s& $ 0-15$ nm & NL \\
Bonaccurso  \cite{bonaccurso02}	& Mica/glass & NaCl solutions & Complete & 1 nm (rms) & $10^2-10^6$ \s & $ 8-9$ nm &L \\
Sun  \cite{sun02}				& Mica/glass& 1-propanol &  $< 90$\d& 1nm (rms) & $10^2-10^6$ \s & $ 10-14$ nm &$-$\\
Bonaccurso  \cite{bonaccurso03}	& Silicon/Glass & Sucrose sol. & Complete & 7 \AA\, (rms) &$10^2-10^6$ \s & $ 0-40$ nm &NL\\
								& Silicon/Glass+KOH &'' &'' & 4 nm (rms) & '' & $ 80$ nm &NL \\
								&''  &'' &'' & 12.1 nm (rms)& ''& $ 100-175$ nm  &NL\\
Neto  \cite{neto03}				& Silica+gold+thiols & Sucrose sol. &$40-70$\d & 6 \AA\,(rms)&$10-10^6$ \s & $ 0-18$ nm &NL\\
Vinogradova  \cite{vinogradova03}	& Silica/glass  & NaCl solutions  &Complete 	&3 \AA (rms) &$10-10^5$  \s &no-slip  &$-$\\
								& Polystyrene & NaCl solutions  &90\d 		&2.5 nm (rms) &'' &  4-10 nm&L\\
Henry  \cite{henry04} 			& Silica/mica & Water & Complete &$-$  & $10^2-10^5$ \s & $ 80-140$ nm & NL\\
 								& Silica/mica+CTAB & CTAB solutions & $>90$\d & $-$ & '' & $ 50-80$ nm &NL\\
Cho  \cite{cho04} 	& Borosilicate+HTS 	& Octane 		&13\d & 3 \AA\,(rms)& $10^2-10^5$ \s& no-slip &$-$\\
					& '' 					& Dodecane 	& 32\d&" &" & no-slip & $-$\\ 
					& '' 					& Tridecane 	& 35\d&" &" & $ 10$ nm & $-$\\ 
					& '' 					& Tetradecane & 37\d&" &" & $ 15$ nm &$-$\\ 
					& '' 					& Pentadecane& 39\d&" &" & $ 10 $ nm &$-$\\ 
					& '' 					& Hexadecane & 39\d&" &" & $ 20 $ nm &$-$\\ 
					& '' 					& Cyclohexane& 25\d&" &" & $ 10$ nm &$-$\\ 
					& '' 					& Benzene 	& 32\d&" &" & $ 50$ nm &$-$\\ 
					& '' 					& Aniline 		& 64\d&" &" & $ 50$ nm &$-$\\ 
					& '' 					& Water 		& 97\d&" &" & $ 30$ nm &$-$\\ 
					& '' 					& Benzaldehyde & 62\d&" &" & $ 20$ nm &$-$\\ 
					& '' 					& Nitrobenzene & 63\d&" &" & $ 10$ nm &$-$\\ 
					& '' 					& 2-nitroanisole & 70\d&" &" & no-slip & $-$\\\ 
\\
\hline
\end{tabular}
\end{center}
\caption{\label{table:AFM} Summary of slip results for experiments using the Atomic Force Microscope (AFM colloidal probe). 
The symbols  used in this table are given in Tables~\ref{table:PD}-\ref{table:SFA}, with additional symbols as:
KOH: potassium hydroxide; 
HTS: hexadecyltrichlorosilane.}
\end{table}


\begin{table}[htdp]\scriptsize
\begin{center}
\begin{tabular}{lllllll}
\hline \\
				& Solid  & Flow & N &   Wetting &  $\displaystyle \frac{k_BT}{\epsilon}$ & Results \\ \\
Koplik  \cite{koplik88}			& HML   & BF & 1536   &  $0- 79$\d & 1.2 & no-slip except at CL\\
Heinbuch   \cite{heinbuch89}		& FL   & BF  & 915 &  Complete &    0.8-2 & $-2\sigma \lesssim \lambda \lesssim 0$ \\
Thompson  \cite{thompson89}		& FL  & CF & 672-5376  & $0-90$\d  & 1.4   & no-slip except at CL \\
Koplik  \cite{koplik89}			& HML   &BF/CF  & 1536-8000  & $0-80$\d  & 1.2   & $\lambda \approx 0 -10$ $ \sigma$ \\
Thompson  \cite{thompson90}		&  HML  &CF  & 672 & $\lesssim 90 $\d  & 1.1   &$\lambda\approx 0-2 $ $\sigma$ \\
Sun  \cite{sun92}				&  HML  & BF  & 7100  & $-$  & 1   & no-slip except for frozen wall\\
Thompson  \cite{thompson97}		&  FL   &  CF& 1152-1728 &   $0-140$\d & 1.1   & $\lambda\approx 0 - 60$ $\sigma$\\
Barrat  \cite{barrat99}			& FL   & BF/CF& 10\,000    &$90-140$\d & 1   &$\lambda\approx 0 -50 $ $\sigma$ \\
Jabbarzadeh  \cite{jabbarzadeh00} 	&  HML   &  CF &  $-$ &   Complete &   9 &$\lambda\approx 0 -10$ nm \\
Cieplak  \cite{cieplak01}			& HML   & BF/CF & $-$  & $-$  &  1.1  & $\lambda \approx 0-15$ $\sigma$ \\
Fan  \cite{fan02}				& HML  & BF &3800 -21090 &Complete   & $-$   & $\lambda \approx 0-5$ $\sigma$\\
Sokhan  \cite{sokhan02}			&  NN  & BF  &  2000& $-$  & $-$   &$\lambda\approx 0-5$ nm \\
Cottin-Bizonne  \cite{cottin-bizonne03} & FL  &  CF & $-$   & $110-137$\d  & 1 &$\lambda\approx 2-57$ $ \sigma$ \\
Galea  \cite{galea04}				& HML   & CF & 6000  & Complete  &  1  & $-3\sigma \lesssim \lambda \lesssim 4$ $\sigma$\\
Nagayama  \cite{nagayama04}		&  FL   &BF  & 2400  & $0-180$\d  & $-$   & $\lambda\approx 0-100$ nm\\ 
Cottin-Bizonne  \cite{cottin-bizonne04} & FL  &  CF & $-$   & $110-137$\d  & 1 &$\lambda\approx 0-150$ $\sigma$ \\ \\
\hline
\end{tabular}
\end{center}
\caption{\label{table:MD}
Summary of MD simulation results for Lennard-Jones liquids with $N$ liquid atoms.  List of symbols:
HML: heavy-mass lattice; 
NN: fixed atoms of carbon nanotube; 
FL:  fixed lattice; 
BF: flow driven by a body force; 
CF: Couette flow; 
CL: contact line. 
}
\end{table}

\section{Experimental methods} 
\label{methods}
As will be discussed below, a large variability exists in the results of slip experiments so it is important to first consider the different experimental methods used to measure slip, directly or indirectly. In these setups, surface conditions may usually be modified by polymer or surfactant adsorption or by chemical modification. Two broad classes of experimental approaches have been used so far, indirect  methods and local methods.

\subsection{Indirect methods}

Indirect methods assume Eq.~\eqref{bc} to hold everywhere in a particular configuration and infer $\lambda$ by measuring a macroscopic quantity. Such methods report therefore effective slip lengths, and they have been the most popular so far. If the effective slip length is $\lambda$, then a system size $L$ at least comparable  $L \sim \lambda$ is necessary in order for slip to have a measurable impact.

\paragraph{Pressure drop versus flow rate.}  

This standard technique is used in many studies \cite{cheng02,breuer03,churaev84,churaev99,schnell56}, where the main results are summarized in Table~\ref{table:PD}\footnote{The dependence of these results on the size of the system was studied in \cite{lauga03}, where there was some evidence that $\lambda$ increased with the size of the system.}. A known pressure drop $\Delta p$ is applied between the two ends of a capillary or a microchannel and the flow rate $Q=\int u \, {\rm d} S$ is measured. A slip boundary condition leads to a  flow rate, $Q(\lambda)$, larger then the no-slip one, $Q_{\rm NS}$, by a factor that varies with the ratio of the slip length to the system size; e.g.,
for a circular pipe of radius $a$, we get
\begin{equation}
\frac{Q(\lambda)}{Q_{\rm NS}}=1+\frac{4\lambda}{a}\cdot
\end{equation}
Using this method, we also note that two groups have reported a larger resistance than that expected with the no-slip condition in microchannels \cite{pfahler90} and for flow through small orifices \cite{hasegawa97}. Their results are not well understood but might be due to electrokinetic effects or flow instabilities.

\paragraph{Drainage versus viscous force.} 
\label{SFA}

This technique consists in imposing the motion (steady or oscillatory) of a curved body perpendicular to a solid surface, and measuring the instantaneous resistive force, which may be compared with that from a model of the fluid motion in the gap, assuming no-slip or slip boundary conditions \cite{happel,persson04,reynolds86}. This method is similar in principle to the pressure drop vs. flow rate method, with the difference that, here, the pressure and velocity fields are unsteady. The two most common narrow-gap geometries are either a sphere of radius $a$ close to a planar surface or two crossed cylinders of radius $a$. For both cases, the viscous force $F$ opposing the motion has the form \cite{vinogradova95}
\begin{equation}\label{drainage_force}
F=-f^*\frac{6\pi \mu a^2 V}{D},
\end{equation}
where $V$ is the instantaneous velocity of the moving body,  $D$ the minimum distance between the two surfaces, and $f^*$ the slip factor. If the no-slip boundary condition is valid, $f^*=1$, otherwise when there is slip, $f^*<1$ and depends on the slip lengths on both surfaces. The calculation for $f^*=f_{\rm slip}$ in the case of equal slip lengths is given by  \cite{hocking73}
\begin{equation}\label{hocking_vino}
f_{\rm slip}=\frac{D}{3\lambda} \left[\left(1+\frac{D}{6\lambda }\right)\ln\left(1+\frac{6\lambda}{D}\right)-1\right],
\end{equation}
and has been extended to account  for two different slip lengths  \cite{vinogradova95} and for the case of any curved bodies  \cite{vinogradova96}. Note that when $D \ll \lambda$, $f_{\rm slip}$ goes to zero as $f_{\rm slip}\sim D\ln(6\lambda/D)/3\lambda$, so that the viscous force, Eq.~\eqref{drainage_force}, only depends logarithmically on $D$; this is a well-known result in the lubrication limit \cite{goldman67}.

Two different experimental apparati have been used to measure drainage forces, the Surface Force Apparatus (SFA) and the Atomic Force Microscope (AFM).  The SFA was invented to measure non-retarded van der Waals forces through a gas, with either a static or dynamic method \cite{israelachvili72,tabor69}, and was extended in \cite{israelachvili78} to measure forces between solid surfaces submerged in liquids. More recently it has been used by many groups to measure slip in liquids, with results summarized in Table~\ref{table:SFA}. This technique usually uses interferometry to report the separation distance between the smooth surfaces. The moving surface is attached to a spring system of known properties so the difference between imposed and observed motions allows a calculation of the instantaneous force acting on the surfaces. 

The AFM was invented  by Binnig, Quate and Gerber \cite{binnig86} and has also been used for many  investigations of slip,  with experimental  results summarized in Table~\ref{table:AFM}. A flexible cantilever beam (typically, microns wide and hundreds of microns long) with a small (tens of microns) attached colloidal sphere is driven close to a surface, either at its resonance frequency or at fixed velocity, and the deflections of the beam are measured. Since the mechanical properties of the beam are known, deflection measurements can be used to infer the instantaneous drainage force on the colloidal particle.

\paragraph{Sedimentation.} 

This experimental method was used in \cite{boehnke99}, with their results summarized in Table~\ref{table:OTHER}. The sedimentation speed under gravity of spherical particles of radius $a$ is measured. If the particles are small enough, their motion will occur at small Reynolds number; in that case, the sedimentation velocity with a slip length $\lambda$, $v(\lambda)$, is larger than its no-slip counterpart, $v_{\rm NS}$, according to
\begin{equation}
\frac{v(\lambda)}{v_{\rm NS}}={\frac{1+3\lambda/a}{1+2\lambda/a}}\cdot
\end{equation}

\paragraph{Streaming potential.} 
\label{streaming}
This is the experimental technique employed in \cite{churaev02}, with their results summarized in Table~\ref{table:OTHER}. A pressure drop is applied to an electrolyte solution between the two ends of a capillary and creates a net flow. Since the surfaces of the capillary acquire in general a net charge in contact with the electrolyte, the net pressure-driven flow creates an advection-of-charges current which results in a surplus of ions on one end of the capillary, and a deficit in the other end. If the two ends of the capillary are not short-circuited, a net steady-state potential difference, termed the streaming potential, exists between the two ends of the capillary and is such that the current due to advection of net charge near the solid surfaces is balanced by the conduction counter-current in the bulk of the electrolyte  \cite{burgeen64,hunter,rice65,saville}. If the fluid experiences slip at the wall (and if the $\zeta$-potential is unchanged by the treatment of the surface), a larger current will occur, hence a large potential difference $\Delta V (\lambda)$ given by
\begin{equation}
\frac{\Delta V (\lambda)}{\Delta V_{NS}}=1+\lambda \kappa,
\end{equation}
where $\kappa$ is the Debye screening parameter, which gives the typical distance close to the surface where there is a net charge density in the liquid, $\kappa ^{-1}= \left({\epsilon_r\epsilon_0 k_BT}/{2e^2n_0}\right)^{1/2}$ \cite{saville}. Here $\epsilon_r$ is the  dielectric constant of the liquid, $\epsilon_0$ the permitivity of vacuum, $k_B$ Boltzmann's constant, $T$ the temperature, $e$ the electron charge and $n_0$ the number density of ions in the bulk of the solution.

\subsection{Local methods} 

All of the methods above have the disadvantage that the slip boundary condition, Eq.~\eqref{bc}, was not verified directly, but instead was estimated via the assumed effect of slip on some other measured macroscopic parameters. A few techniques have been introduced that try to alleviate this indirect estimation of slip.

\paragraph{Particle image velocimetry (PIV).} 

This method was proposed to investigate slip in \cite{joseph05,meinhart02,meinhart04}; we have summarized their results in Table~\ref{table:OTHER} (see also  \cite{jin04,yamada99,zettner03}). Let us consider a pressure-driven flow between two parallel plates with  separation distance $2h$. In this case, a non-zero slip length $\lambda$ leads to a 
velocity field
\begin{equation}
U_{\rm slip}(z)= -\frac{h^2}{2\mu}\frac{{\rm d} p}{{\rm d} x}\left[1-\frac{z^2}{h^2}  +\frac{2\lambda}{h}\right],
\end{equation}
which shows that a change in the condition at the boundary has a bulk effect: the (no-slip) Poiseuille flow is augmented by a plug flow. The idea of PIV is then to use small particles as passive tracers in the flow,  to measure the velocities of the particles with an optical method and check whether the velocities extrapolate to zero at the solid surface. Since small particles have large diffusivities, results need to be averaged to extract the advective part of the tracer motion. Also, the particles in general move relative to the fluid owing to hydrodynamic interactions \cite{happel}, so care is needed when interpreting the measured velocities.

\paragraph{Near-field laser velocimetry using fluorescence recovery.} 

This is the experimental technique proposed in \cite{leger03,pit99,pit00}, and we have summarized their results in Table~\ref{table:OTHER}. In this method, the velocity field of small fluorescent probes is measured close to a nearby surface. An intense laser illuminates the probes and renders them non fluorescent (photobleaching). Monitoring the fluorescence intensity in time  using evanescent optical waves allows to obtain an estimate of the slip length. Note that the fluorescence intensity evolves in time due to both convection (part which depends on slip) and molecular diffusion, so a careful analysis is needed. Also, because of the fast diffusion of molecular probes, the method is effectively averaging over a diffusion length $\sim1$ $\mu$m, which is much larger than the  evanescent wavelength.

\paragraph{Fluorescence cross-correlations.} 

This is the latest experimental method, proposed by \cite{lumma03}, and their results are also summarized in Table~\ref{table:OTHER}. Fluorescent probes excited by two similar laser foci are monitored in two small sample volumes separated by a short distance. Cross-correlation of the fluorescence intensity  fluctuations due to probes entering and leaving the observation windows allows to determine both the flow direction and intensity. The measured velocities are averaged over the focal size of microscope and the characteristics of the excitation laser.

\section{Molecular dynamics simulations} 
\label{MD}
Molecular Dynamics (MD) simulations are useful theoretical tools in the study of liquids \cite{allen87,koplik95_annurev,muser00} which have been extensively used to probe boundary conditions. We summarize below the principle of the technique and discuss the interpretation of their results in the continuum limit.

\subsection{Principle}
MD simulations integrate numerically Newton's law of motion  for single atoms (or molecules)
\begin{equation}\label{Newton}
m_i\frac{{\rm d}^2{\bf r}_i}{{\rm d} t^2} = \sum_{j}{\bf F}_{ij},
\end{equation}
where  $m_i$ is the atomic mass, ${\bf r}_i$ the position of atom $i$, and ${\bf F}_{ij}$ the interatomic (or intermolecular) force between atoms $i$ and $j$, that is ${\bf F}_{ij}=-\nabla_i V_{ij}$ where $V_{ij}$ is the interaction potential. Potentials used in simulations range from the Lennard-Jones two-body potential
\begin{equation}\label{LJ}
V_{ij}=\epsilon\left[\left(\frac{\sigma}{r_{ij}}\right)^{12}-c_{ij}\left( \frac{\sigma}{r_{ij}}\right)^6 \right],
\end{equation}
where $\epsilon$ is an energy scale, $\sigma$ the atomic size, and $r_{ij}$ the distance between atoms $i$ and $j$, to more realistic potentials including many-body or  orientation-dependent interactions \cite{allen87,koplik95_annurev,muser00}. The set of Eqs.~\eqref{Newton} are integrated in time, with appropriate numerical cut-offs, and with specified boundary conditions and initial conditions. Usually initial positions are random and initial velocities are taken from a Boltzmann distribution.  It is also possible to modify Eq.~\eqref{Newton} slightly to model evolution at constant temperature either by coupling the system of atoms to a heat bath or by a  proper rescaling of the velocities at each time step. Interactions with a solid can occur by adding different wall atoms, either fixed on a lattice or coupled to a lattice with a large spring constant, to allow momentum transfer from the liquid but prevent melting. The constants $(c_{ij})$ in Eq.~\eqref{LJ} allow variation of the relative intermolecular attraction between liquids and solids, which therefore mimicks wetting behavior. Using a simple additive model \cite{israelachvili_book}, the case of a partially wetting fluid with contact angle $\theta_c$ can be modeled with
\begin{equation}
\cos\theta_c = -1 +2\frac{\rho_{\rm S}c_{\rm LS}}{\rho_{\rm L}c_{\rm LL}}
\end{equation}
where $\rho_{\rm S}$ ($\rho_{\rm L}$) is the solid (liquid) density and $c_{\rm LS}$ and $c_{\rm LL}$ are, respectively, the liquid-solid and liquid-liquid intermolecular constants. Finally, two types of flow can be driven. In the first kind, atoms that constitute the wall(s) are driven at a constant velocity and the bulk liquid has a Couette-flow profile. In the second kind, each liquid atom is acted upon by a body force and the liquid has a Poiseuille-flow profile.

\subsection{Results}

The method described above has been used  to study slip in different types of liquids 
\cite{barrat99,cieplak01,cottin-bizonne04,cottin-bizonne03,fan02,galea04,heinbuch89,jabbarzadeh00,koplik88,koplik89,nagayama04,sokhan02,sun92,thompson89,thompson90,thompson97}, with results summarized in Table~\ref{table:MD}.  Early simulations showed no-slip except near contact lines \cite{koplik88,thompson89}. More recent investigations have reported that  molecular slip increases with decreasing liquid-solid interactions \cite{barrat99,cieplak01,nagayama04}, liquid density \cite{koplik89,thompson90}, density of the wall \cite{thompson97}, and decreases with pressure \cite{barrat99}.  The model for the solid wall, the wall-fluid commensurability and the molecular roughness were also found to strongly influence slip \cite{barrat99,bocquetbarrat93,galea04,jabbarzadeh00,sun92,thompson97}. Related investigations include contact line motion \cite{freund03} (and references therein), motion of a sliding plate \cite{koplik95} and the validity of Stokes drag formulae at small scales \cite{vergeles96}.

\subsection{Interpretation in the continuum limit}\label{issues}

Results from MD simulations can sometimes be difficult to interpret in the continuum limit. First, for computational reasons, simulations to date are limited to tens of thousands of atoms, which restricts the size and time scale of the simulated physical system. The three control parameters in the simulations are the molecular/atomic mass $m$, the interaction energy $\epsilon$, and the molecular/atomic size $\sigma$. Consequently, lengths are measured in units of  $\sigma$ ($\sim 3$ \AA\,) and times in units of  the molecular time scale  $\tau\sim \sqrt{m\sigma^2/\epsilon}$ ($\sim 10^{-12}$~s). Simulated systems are therefore limited to tens of nanometers, and times scales to nanoseconds. The consequence of this observation is that MD simulations always probe systems with much higher shear rates than any experimental setup.  For example, in MD simulations of Couette flow, the typical wall velocity is $U\sim \sigma/\tau$, corresponding to typical shear rates $\g\sim \sigma/\tau h$ where $h$ is the typical length scale of the simulation box, usually a few tens of $\sigma$. Consequently,  $\g\sim 10^{11}$ \s, which is orders of magnitude larger than experimental shear rates. Note that this does not apply to investigations inferring slip length from equilibrium simulations \cite{bocquetbarrat93,bocquet94}.

A second significant issue in interpreting results of MD simulation was pointed out by Brenner and Ganesan in the case of particle diffusion near a solid surface \cite{brenner00}, and concerns the scale separation between molecular and continuum phenomena. The idea is that the correct boundary condition in the continuum realm should arise asymptotically as a matching procedure between the outer limit of the inner (molecular) system and the inner limit of the outer (continuum) system.
By doing so, the change in the physical behavior within a few intermolecular length scales of the surface is explicitly taken into account, which allows to make a distinction between {\it conditions at a boundary} and {\it boundary conditions}. As a consequence, slip lengths should not be measured literally at the molecular scale but arise as the  extrapolation,  at the boundaries, of the far field hydrodynamic results, a procedure which is not always performed appropriately.

\section{Discussion: Dependence on physical parameters} 
\label{discussion}

Having described the different methods by which slip is investigated, we present in this section a discussion of both experimental and simulation results and compare them with theoretical models. The discussion is organized according to the physical parameters upon which slip has been found to depend.

\subsection{Surface roughness} 

\paragraph{Roughness influences resistance.}

Be it at the molecular size \cite{galea04} or on larger scales \cite{bonaccurso03,georges93, jabbarzadeh00,pit99,granick02}, roughness and geometrical features have been observed to influence the behavior at liquid-solid interfaces. Not only does roughness leads to an ambiguity as to the exact location of the surface, but it impacts the dynamics of the nearby fluid, leading experimentally either to an increase  \cite{georges93,jabbarzadeh00,pit99,granick02} or a decrease  \cite{bonaccurso03} of friction with roughness.

\paragraph{Roughness decreases slip.}

The physical idea for a roughness-induced resistance is straightforward: on the roughness length scale, a flow is induced that dissipates mechanical energy and therefore resists motion. For the same reason, a bubble with a local no-shear surface rises at a finite velocity in a liquid. More generally, geometrical features of size $a$ on a surface can be solely responsible for a large resistance on large scales $L\gg a$, independently of the details of the local boundary condition on the scale $a$. This feature was first recognized by Richardson \cite{richardson73} who assumed a periodic perfectly-slipping surface shape and performed asymptotic calculations for the limit $a/L \to 0$; in this limit the no-slip boundary condition was recovered (see also \cite{nye69,nye70}). The calculation was revisited by Jansons  \cite{jansons88} who considered a  small fraction $c$ of roughness elements of size $a$ with a local no-shear condition on an otherwise perfectly slipping surface. Equating the viscous force  associated with the disturbance flow created by the defects, ${\cal O}(\mu\g d^2)$, to the local Stokes drag on a defect, ${\cal O}(\mu a u_s)$, where $d\sim a/c^{1/2}$ is the typical distance between defects and $u_s$ is the fluid velocity near the defects,  leads to an effective slip length for the surface,  $\lambda=u_s/{\dot \gamma}$, given by
\begin{equation}
\lambda \sim \frac{a}{c}\cdot
\end{equation}
When $c$ is of order one, all these length scales $(a,d,\lambda)$ are of the same order and results of \cite{richardson73} are recovered. Recently more rigorous results were derived in  \cite{casado-diaz03}.

When the boundary condition is locally that of no-slip, roughness shifts the position of the effective surface into the liquid. Calculations have been made for periodic and random surfaces \cite{miksis94,ponomarev03,sarkar96,tuck95} and are related to earlier work on the boundary conditions for porous materials \cite{richardson71,taylor71} and the Laplace equation \cite{sarkar95}.

\paragraph{Roughness-induced dewetting.}

The interaction of roughness with surface energies can lead to the spontaneous dewetting of a surface and the appearance of a super-hydrophobic state, as proposed in \cite{cottin-bizonne04,cottin-bizonne03}. In that case, roughness could increase slip by producing regions of gas-liquid interface at the solid boundary. Let us consider for illustration a surface $S$ covered with a fraction $c$ of roughness elements of height $a$ in a liquid at pressure $p$ and let us denote by $r>1$ the ratio of real to apparent surface area. In that case, the change in free energy, $\Delta G$, to dewet the apparent area $(1-c)S$  between the roughness elements arises from surface energies and work done against the liquid
\begin{equation}
\Delta G = r(1-c)S(\gamma_S-\gamma_{LS}) + (1-c)S(pa+\gamma),
\end{equation}
where $\gamma$ is the surface tension of the liquid, $\gamma_S$ that of the solid and $\gamma_{LS}$ the  liquid-solid interfacial tension. Consequently, 
using Young's law, $\gamma \cos\theta_c=\gamma_S-\gamma_{LS}$, we see that dewetting is energetically favorable when $p<-{\gamma}(1+r\cos\theta_c)/a$, which, for a given value of the pressure, will occur if the surface is hydrophobic $(\cos\theta_c<0)$ and $a$ is small enough. The super-hydrophobic state is therefore due to a combination of geometry and wetting characteristics. 

This idea is related to the so-called fakir droplets  \cite{bico99,cassie44,onda96,wenzel36} and to more general drag reduction mechanisms found in nature using gas bubbles \cite{buschnell91}. Trapped bubbles in rough surfaces were studied by \cite{hocking76} in the context of contact line motion and are  probably responsible for the apparent slip lengths reported in  \cite{watanabe98,watanabe99,watanabe98_2} for flow over fractal surfaces, and possibly other studies as well (see also \cite{lauga03}). A similar  mechanism was quantified experimentally using trapped bubbles in rough silicon wafers  \cite{ou04} (see also the calculations in  \cite{wang03}) and show promise of decreasing turbulent skin-friction drag \cite{min04}.

\subsection{Dissolved gas and bubbles} 

\paragraph{Slip depends on dissolved gas.}

The amount of slip has been observed experimentally to depend on the type and quantity of dissolved gas.  It is reported in  sedimentation studies  \cite{boehnke99} that slip was not observed in vacuum conditions but only when the liquid sample was in contact with air.  Furthermore, the study in \cite{granick03} showed that tetradecane saturated with CO$_2$ lead to results consistent with no-slip but significant slip when saturated with argon, whereas the opposite behavior was observed for water. Similar results were reported in \cite{tretheway04}. More generally, slip results in non-wetting systems are found to  depend strongly  on the environment in which the experiment is performed \cite{cottin-bizonne05}.

\paragraph{Flow over gas: Apparent slip.}

The results above, together with experiments showing dependence of slip on the absolute value of the pressure \cite{tretheway04}, and spatially-varying velocity fields  \cite{meinhart04}, hint at the possibility of flow over surface-attached gas pockets or bubbles (see also the discussion in  \cite{lauga03}). Recent results in \cite{cottin-bizonne05} also point at the possibility of flow over gas pockets associated with the contamination of hydrophobic surfaces by nanoparticles. We also note that the group of Steve Granick reported a contamination of their previous ostensibly smooth mica surfaces by platinum nanoparticles \cite{lin03}, possibly affecting some of their experimental results in \cite{granick01,granick02_macro,granick02,granick02_langmuir}.

The idea of a flow over a gas layer was first mentioned in \cite{ruckenstein83} and revisited in  \cite{ruckenstein91} as a possible explanation for the attraction between hydrophobic surfaces in water: the attraction could be due to the hydrodynamic correlated fluctuations of the gas interfaces, analogous to the Bjerknes force between two pulsating bubbles. Detailed theoretical considerations have shown that it would be favorable for water between two hydrophobic surfaces to vaporize \cite{lum99}. Flow of binary mixtures have also been shown to phase separate by the sole action of intermolecular forces \cite{andrienko03}.

It is clear that flow over a layer of gas will lead to an apparent slip. Since stress must be continuous at a liquid-gas interface, a difference of shear viscosities will lead to a difference of strain rates. If a liquid of viscosity $\mu_1$ flows over a layer of height $h$ with viscosity $\mu_2$, the apparent slip length for the flow above is  (see, e.g., \cite{vinogradova95})
\begin{equation}\label{2layer}
\frac{\lambda}{h}=\frac{\mu_1}{\mu_2}-1;
\end{equation}
 ${\mu_1}/{\mu_2}\approx 50$ for a gas-water interface.  Three differences exist however between a flow over a gas layer and flow over a set of bubbles: (a) The gas in bubbles recirculates, which decreases the previous estimate Eq.~\eqref{2layer} by about a factor of four; (b) No-slip regions located between the bubbles will also significantly decrease the apparent slip lengths \cite{jansons88,lauga03,philip72,philip72_2,richardson73} (see also \cite{alexeyev96} on the effect of non-uniform slip lengths); (c) Bubbles are in general not flat, which decreases the previous estimates even further (similar to the effect of roughness on a shear flow).
 
When the gas layer is in the Knudsen regime ($\sigma \ll h \ll \ell_m$), the shear stress in the liquid, ${\cal O}(\mu \g)$, is balanced by a purely thermal stress in the gas, ${\cal O}( \rho u_s u_{\rm th})$, where $\rho$ is the gas density, $u_{\rm th}$ the thermal velocity $u_{\rm th}={\cal O}( \sqrt{k_BT/m})$ ($m$ is the mass of a gas molecule) and $u_s$ is the liquid velocity at the interface. Balancing these two contributions leads to an apparent slip length, $\lambda = u_s/\dot{\gamma}$, given by \cite{degennes02}
\begin{equation}\label{degennes_gas}
\lambda \sim \frac{\mu}{ \rho u_{\rm th}},
\end{equation}
which is independent of $h$ and can be as large as microns. Note that the slip length given by Eq.~\eqref{degennes_gas} increases with the viscosity of the liquid.

\paragraph{Nanobubbles in polar liquids?}

Over the last four years, many groups have reported experimental observation of nanobubbles against hydrophobic surfaces in water \cite{attard02,holmberg03,ishida00,lou02,simonsen04,steitz03,switkes04,tyrrell01,tyrrell02,zhang04,zhang04_2}, with typical sizes  $\sim 10-100$ nm and large surface coverage (see also the  reflectivity measurements in \cite{jensen03,schwendel03}). The nanobubbles disappear when the liquid is degassed. Similar bubbles could be responsible for slip measurement in some of the experiments to date (see also \cite{vinogradova01,yakubov00}). How could the formation of such bubbles be explained? Thermal fluctuations lead to bubble sizes $a\sim \sqrt{k_BT/\gamma}$, which are on the order of the molecular length. It has been proposed that shear might induce bubbles, but the mechanism is not clear \cite{degennes02}. An alternative scenario could be a local decrease in pressure near hydrophobic surfaces due to intermolecular forces  (see, e.g.,  Eq. 4.23 in \cite{degennes85}).

The second important issue for nanobubbles is their stability against dissolution. 
A spherical gas bubble of radius $a$, diffuses into the liquid on a timescale  \cite{epstein50,ljunggren97}
\begin{equation}
\tau\sim \frac{Mp_0a^2}{Dc_0RT}\left(1+\frac{p_0a}{\gamma} \right)
\end{equation}
where $M$ is the molar mass, $p_0$ the far-field pressure, $D$ the diffusion coefficient of the gas in the liquid, $T$ the temperature, $c_0$ the saturated gas concentration in the liquid (mass per unit volume) at pressure $p_0$,  $\gamma$ the liquid surface tension  and $R$ the absolute gas constant; note that this estimate is independent of surface tension for sufficiently small bubbles. For a 10 nm bubble, $\tau \approx 10$ $\mu$s and $\tau$ becomes a few hours  when $a\approx 100$ $\mu$m. It has therefore been argued that the existence of such small bubbles can only be explained if the liquid is supersaturated with gas \cite{attard02}. In many pressure-driven flow experiments at small scale, a high-pressure gas in contact with the liquid is used to induce the flow; for example pressured gas at 10 atm is used to drive motion in \cite{breuer03}, equivalent to the internal pressure of a 100 nm bubble. However, the resulting equilibrium between a gas bubble and a supersaturated solution is well-known to be unstable as any perturbation either grows without limit or dissolves away. A possible resolution to the stability problem of such bubbles might come from intermolecular forces in the gas which become important when bubbles reach  small radii and large pressures \cite{wentzell86}.

\subsection{Wetting} 

\begin{figure}[t]
\centering
\includegraphics[width=.8\textwidth]{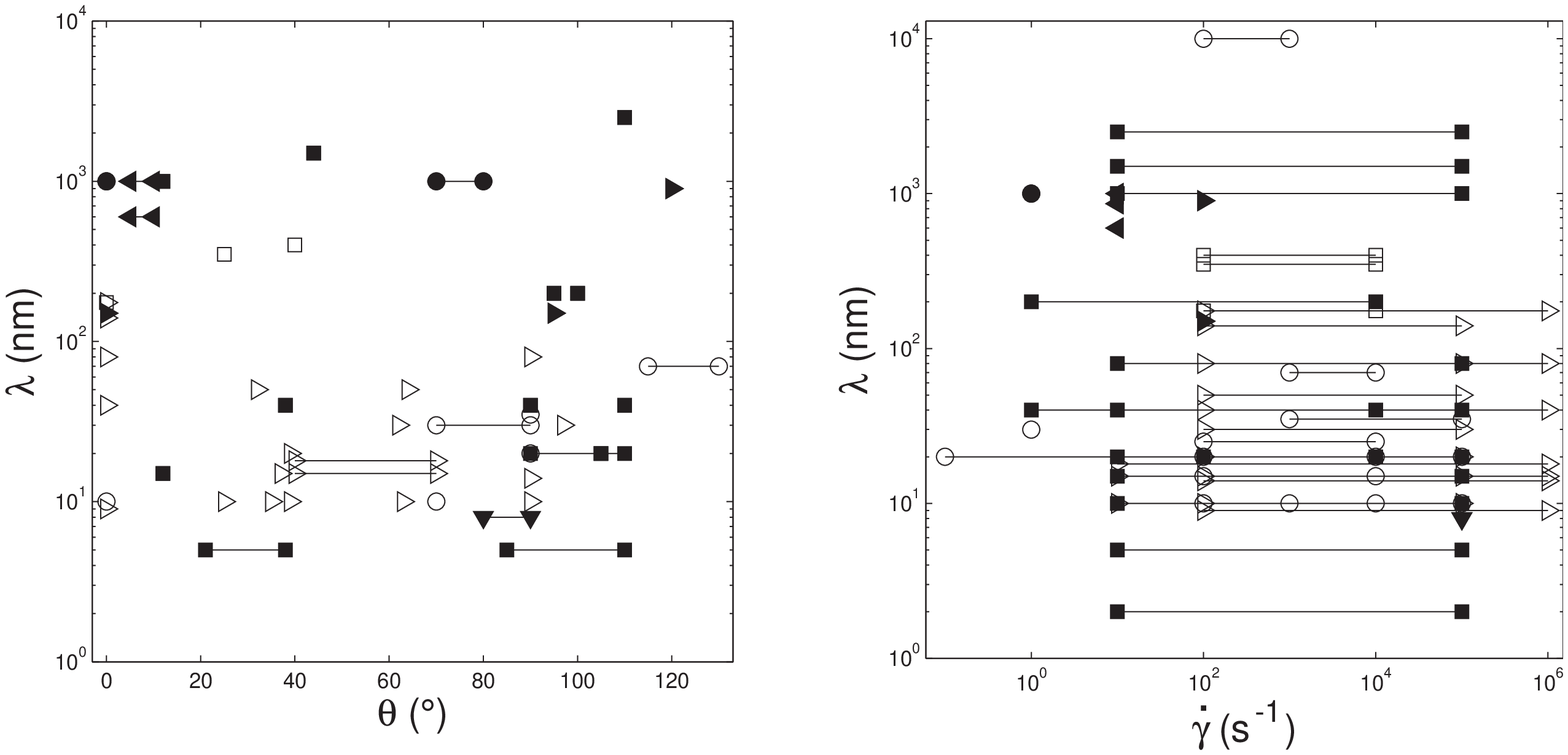}
\caption{\label{slip_depends}  Experimental variation of the slip length, $\lambda$, with the liquid-solid contact angle, $\theta$ (left), and the typical experimental shear rate, $\dot{\gamma}$ (right), for the experimental results summarized in Tables~\ref{table:PD}-\ref{table:AFM}: 
Pressure-driven flow ($\circ$), 
sedimentation ($\bullet$), 
fluorescence recovery ($\square$), 
PIV ($\blacktriangleright$), 
streaming potential ($\blacktriangledown$), 
fluorescence cross-correlations  ($\blacktriangleleft$), 
SFA ($\blacksquare$), and
AFM ($\triangleright$). When a solid line is drawn, the experimental results are given for a range of contact angles and/or  shear rates. Furthermore, when the value of the contact angle is unknown, the results are not reported.}
\end{figure}

\paragraph{Slip depends on wetting properties.}

It was recognized early that friction at the liquid-solid boundary should be a function of the physicochemical nature of both the solid and the liquid \cite{goldstein}. In particular, the wetting properties have been found to play a crucial role in many experiments. Wetting of solids by liquids is reviewed in \cite{degennes85,degennes_book} and is quantified by the spreading coefficient, $S= \gamma_S-\gamma- \gamma_{LS}$, which is the difference in surface energy between a dry solid surface and the same surface wet by a liquid layer ($\gamma_S$, $\gamma$ and $\gamma_{LS}$ are the solid, liquid and liquid-solid interfacial energies). When $S>0$, the solid is completely wet by the liquid and when $S<0$ the wetting is partial. In the latter case, a small liquid droplet on the solid surface would take the shape of a spherical cap with a contact line at an angle $\theta_c$ to the solid, where $\theta_c$ is the equilibrium contact angle, and is give by Young's law, $\gamma \cos\theta_c=\gamma_S-\gamma_{LS}$.
The surface is said to be hydrophobic if $\theta_c>90$\d, and in that case the nucleation of small bubbles in the liquid should occur preferentially on the surface.

Slip has been measured for systems in complete wetting \cite{bonaccurso03,bonaccurso02,henry04,leger03,pit99,pit00} and
partial wetting \cite{baudry01,boehnke99,cheikh03,cho04,breuer03,churaev02,churaev84,cottin-bizonne05,cottin02,craig01,henry04,joseph05,churaev99,leger03,lumma03,neto03,sun02,meinhart02,meinhart04,vinogradova03,granick01,granick02,granick02_langmuir}.
The amount of slip has usually been found to increase with contact angle, either systematically (e.g., \cite{granick01}) or only for non-polar liquids \cite{cho04}.  All these results are summarized in the plot displayed in Fig.~\ref{slip_depends} (left), which shows however, overall, a poor correlation between slip and contact angle.

\paragraph{The Tolstoi theory.}

It appears that Tolstoi was the first to try to quantify the importance of surface energies on slip at the molecular level \cite{blake90,frenkel55,tolstoi52}. Using concepts from macroscopic thermodynamics at the molecular scale, Tolstoi considered the relation between surface energies and molecular mobility (hence diffusivity) near a solid surface by calculating the work it takes for molecules to make room for themselves  in the liquid, and how that changes near a boundary. The molecular diffusivity $D$ is given by the product of the molecular scale, $\sigma$, and a velocity, $D\sim \sigma V$. The molecular velocity is $V\sim\sigma/\tau$, where $\tau$ is the typical time scale for hopping from one molecular position to the other. This is typically an inverse molecular frequency corrected for the energy it takes to create a void of size $\sigma$, which is  similar to the surface energy $\gamma \sigma^2$. Near a solid, this energy involves the solid and liquid/solid interfacial energies, hence the possibility of having a different molecular mobility close to a surface. In the case of complete wetting, the Tolstoi model leads to the no-slip boundary condition within $\pm$ one molecular layer in the liquid, but in the case of partial wetting, molecules near the surfaces are found to have larger mobilities, leading to a slip length \cite{blake90,tolstoi52}
\begin{equation}\label{tolstoi}
\frac{\lambda}{\sigma}\sim \exp\left({\displaystyle \frac{\alpha\sigma^2\gamma(1-\cos\theta_c)}{k_B T}}\right)-1,
\end{equation}
where $\alpha$ is a dimensionless geometrical parameter of order one, $\gamma$ is the liquid surface tension and $\theta_c$ the equilibrium contact angle. The estimate, Eq.~\eqref{tolstoi}, increases with the contact angle and can be orders of magnitude above the molecular length.

\paragraph{Intrinsic (molecular) slip.}

Another theory at the molecular scale uses the  fluctuation-dissipation theorem and Green-Kubo relations to derive slip lengths from equilibrium thermodynamics \cite{barratbocquet99,bocquetbarrat93,bocquet94}. Using Onsager's hypothesis of linear regression of fluctuations, {\it i.e.}, that small fluctuations around equilibrium can be described by the same equations that describe the relaxation from non-equilibrium,  leads to a formula for the time-dependent momentum correlation function in the liquid, function of both the slip length and the wall position. The boundary condition is found to be applied about one molecular layer inside the liquid and the slip length is given by
\begin{equation}
\frac{\lambda}{\sigma}\sim \frac{D^*}{S_tc_{\rm LS}^2\rho_c \sigma^3},
\end{equation}
where $D^*=D_\parallel / D_0$, $D_\parallel$ is the collective molecular diffusion coefficient, $D_0$ the bulk diffusivity, $S_t$ the structure factor for first molecular layer (both $D^*$ and $S_t$ are dimensionless numbers of order unity), $\rho_c$ the fluid density at the first molecular layer (number per unit volume), and $c_{\rm LS}$ the dimensionless liquid-solid coefficient of the Lennard-Jones potential, Eq.~\eqref{LJ}. In the case of complete wetting, the slip length is essentially zero as soon as the roughness is a few percent of the molecular size, but in a non-wetting situation the slip length  can be up to two orders of magnitude above molecular size and increases with contact angle \cite{barratbocquet99}.  As the macroscopic contact angle goes to 180\d, the slip length diverges as $\lambda/\sigma\sim 1/(\pi-\theta_c)^4$. The theoretical predictions are found to agree very well with MD simulations \cite{barratbocquet99,barrat99} and the  approach was extended to polymer solutions in \cite{priezjev04}.
The details of the molecular slip mechanism were subsequently studied in \cite{lichter04} which showed that slip at low shear rates occurs by localized defect propagation and switches to slipping of whole molecular layers for larger rates.

\subsection{Shear rate} 

\paragraph{Slip depends on shear rates.}

In many investigations, slip was observed to depend on the shear rate at which the experiment or simulation was performed (see Fig.~\ref{slip_depends}, right). When that is the case, the slip boundary condition, Eq.~\eqref{bc}, becomes nonlinear $\lambda=\lambda(\g)$, and we refer to this situation as NL in Tables~\ref{table:PD}-\ref{table:AFM}. Shear-dependent slip was reported experimentally in \cite{bonaccurso03,breuer03,churaev84,craig01,henry04,neto03,granick01,granick02_macro,granick02,granick02_langmuir}, with the strongest dependence to date in \cite{granick01}. When such a dependence is not observed, Eq.~\eqref{bc} is linear (L) and the slip length is a property of the liquid-solid pair. Two drainage experiments have also reported linear boundary conditions, with  force proportional to velocity in Eq.~\eqref{drainage_force}, but with a profile, for small separations between the surfaces, which differs from that given by a uniform slip length model  \cite{baudry01,cottin02}.

Most MD simulations report boundary conditions in the linear regime, except in \cite{heinbuch89,nagayama04}, where the magnitude of slip was found to depend on the magnitude of the driving force, and in \cite{thompson97} where simulations give slip lengths that diverge at high shear rates  $\lambda/\sigma\sim (1-\g/\g_c)^{-1/2}$. 

\paragraph{The leaking mattress model.}

A mechanical model for shear-dependent  slip in drainage experiments was proposed in \cite{laugabrennerslip04}, based on the assumption that a layer of gas bubbles is present on the solid surface. Since drainage experiments are unsteady, the bubble sizes will be a function of time in response to pressure variations in the liquid (by the combination of compression and diffusion), hereby modifying the amount of liquid which is necessary to drain out at each cycle of the oscillation, and therefore modifying the viscous force on the oscillating surface. This idea of flow over a ``leaking mattress'' leads to a frequency-dependent decrease in the viscous force,  $f^*(\omega)$ in Eq.~\eqref{drainage_force}, given by
\begin{equation}\label{mattress}
\frac{f^*(\omega)}{f_{\rm slip}}= \frac{1}{1+\left(\delta k_1
+\displaystyle \frac{( \omega \delta k_2)^2}{1+\delta
k_1}\right)},\quad k_1 = \frac{n a_0^2I(\theta_c)c_0\sqrt{ \omega \kappa}}{\pi \rho_0p_0 D a},\quad
k_2 =  \frac{c_0 h_0}{\pi c_{\infty} p_0  D a}\left(1+\frac{ c_{\infty}(D-2h_0)
}{\rho_0h_0}\right),
\end{equation}
where $f_{\rm slip}$ is the slip factor due to flow over bubbles, given by Eq.~\eqref{hocking_vino} ({\it i.e.}, $f_{\rm slip}$ is the zero-frequency force decrease, associated with  a slip length, $\lambda$, which describes the effective resistance of the covered solid surface), and the other terms quantify the dynamic response of bubbles and contribute to an additional force decrease. In Eq.~\eqref{mattress},  $a$ is the curvature of the surface, $c_{\infty}$ is the far field dissolved gas concentration (mass per unit volume), $p_0$ is the far-field liquid pressure, $c_0$ the dissolved gas concentration in equilibrium with gas at pressure $p_0$, $\rho_0$ the gas density at pressure $p_0$, $\omega$ the frequency of oscillation of the drainage experiment, $D$ the minimum distance between the two surfaces, $\mu$ the liquid shear viscosity, $n$ the number of bubbles per unit area, $a_0$ the equilibrium radius of curvature of the bubbles, $\kappa$ the diffusivity of the gas in the liquid, $I$ a geometrical shape factor of order unity, $h_0$ the mean bubble height on the surface and $\delta={12\pi\mu a^2f_{\rm slip}}/{D}$. Note that $f^*(\omega)$ increases with the liquid viscosity and the curvature of the surfaces. The results of this model  compare well with the experiments of \cite{granick01}.

\paragraph{The critical shear-rate model.}

An empirical model for shear-dependence inspired by the data in \cite{granick01} was also proposed in \cite{spikes03}. Slip is assumed to occur locally with a constant slip length $\lambda$ as soon as the local shear rate reaches a critical value $\g_c$; below this critical value, the no-slip boundary condition is assumed to remain valid. This model has therefore slip confined to an annular region around the narrow gap where shear rates are the highest. With the two fitting parameters $(\lambda,\g_c)$, the model can reconcile various experimental data, except for small separations of the surfaces.

\paragraph{Viscous heating.}

In a steady flow, the rate of dissipation of mechanical energy (which, for a Newtonian fluid, depends on the square of the shear rate) is equal to the rate of change of internal energy due to changes of temperature. Since the viscosity depends on temperature, high shear rates could lead to the possibility of viscous heating \cite{Gavis},  a flow-induced reduction in viscosity, which could be interpreted as an apparent slip. Assuming a traditional exponential law for the viscosity $\mu=\mu_0\exp[{-\beta (T-T_0)/T_0}]$ and flow in a circular capillary of radius $a$, the apparent slip length due to viscous heating would be \cite{lauga03}
\begin{equation}
\frac{\lambda}{a}\sim \frac{\beta}{T_0}\left(\frac{\nu}{\kappa_T}\right)\frac{(\dot{\gamma}a)^2}{c_p},
\label{temp}
\end{equation}
where $T_0$ is the reference temperature, $\beta$ a dimensionless  coefficient of order one, $\nu$ the fluid kinematic viscosity, $\kappa_T$ the fluid thermal diffusivity and $c_p$ the specific heat. Although viscous heating can be neglected in most experiments to date, it has the potential of becoming important at  higher shear rates (see also \cite{urbanek93} on the issue of temperature variations).

\subsection{Electrical properties} 

\paragraph{Apparent slip depends on ionic strength and polarity.}

When probing slip in electrolyte solutions and polar liquids, the amount of slip was found to vary with electrical properties. Sedimentation experiments reported that slip was only observed for polar liquids \cite{boehnke99}. Fluorescence-correlation measurements reported slip lengths of the same order as the screening length $\kappa^{-1}$, which decrease with the ionic strength of the solution \cite{lumma03}. Drainage experiments also reported that when liquids are polar, slip fails to increase with hydrophobicity but increases with the dipolar moment of the liquid \cite{cho04}; this result was interpreted as a disruption  by the drainage flow of the local liquid cohesive energy arising from dipole-image dipole interactions close to the surface \cite{cho04}. Finally, we note that the morphology of nanobubbles has also been observed to depend on pH  \cite{tyrrell02} (see also the discussion in \cite{bunkin97}).

\paragraph{Electrostatic-induced averaging.}

When using small tracers to probe the fluid velocity, electrical effects need to be carefully taken into account. The first issue concerns measurement close to the surface. If the surface and the particles are similarly charged, particles will be repelled electrostatically  and will not come within a distance $\sim \kappa^{-1}$ of the surface (if charges are opposite, tracers will stick to the the surface). This effect will therefore increase the average velocity of the particles when compared to what would be expected if electrical effects were not considered. If the averaging window has a height $h > \kappa^{-1}$ above the surface, then the mean flow velocity will appear increased by a factor $(1+1/\kappa h)$, which, if interpreted as a slip length would give an apparent value
\begin{equation}
\frac{\lambda}{h}=\frac{1}{\kappa h -1}\cdot
\end{equation}

\paragraph{Apparent slip due to charged tracer.}

The other potential problem with experimental methods using small tracers is the influence of the streaming potential  on their motion \cite{lauga04}. If the particles are charged, their velocity will also include an electrophoretic component in response to the flow-induced potential difference; moreover, if they are sufficiently charged, this velocity will be able to overcome the electro-osmotic back-flow  in the bulk and the particles will move faster than the local liquid, leading to an apparent slip length. If we consider the case of a pressure driven flow between two parallel plates separated by a distance $2h$, the resulting apparent slip length is given by \cite{lauga04}
\begin{equation}
\frac{\lambda}{h} \sim  \frac{\zeta_w(q\zeta_p-\zeta_w)(\epsilon_r\epsilon_0)^2}
{2\sigma_e\mu h^2+ (\epsilon_r\epsilon_0\zeta_w)^2\kappa h},
\end{equation}
where $\zeta_w$ is the wall zeta-potential, $\zeta_p$ the particle zeta-potential, $q$ a dimensionless factor of order one depending on the ratio of the screening length to the particle size, $\mu$ the shear viscosity of the liquid, $\sigma_e$ the electrical conductivity of the liquid and $\epsilon_r$ its  dielectric constant. In the case of low-conductivity electrolytes, such apparent slip length can be as large as hundreds of nanometers.

\subsection{Pressure} 

\paragraph{Apparent slip depends on pressure.}
In the velocimetry experiments of \cite{tretheway04}, the measured slip length was found to decrease with the value of the absolute pressure; for water, the no-slip boundary condition was recovered when the absolute pressure reached 6 atm. Such results are another hint at the likely role of surface-attached bubbles, presumably decreasing in size with an increase in pressure.

\paragraph{Slip due to pressure gradients.}
The possibility of surface slip due to gradients in liquid pressure was proposed in \cite{ruckenstein83} using arguments from equilibrium thermodynamics. The idea is that the chemical potential of a liquid molecule depends on pressure, so a pressure gradient leads to a gradient in chemical potential, hence a net force $F$ on the liquid. Assuming $ F\ll k_B T/ \sigma$, a molecular model of diffusion under force allows us to get the net surface velocity and estimate the slip length. For a circular pipe of radius $a$, the slip length is predicted to be 
\begin{equation}\label{rucken}
\lambda \sim \frac{\mu D_S}{\rho a k_BT}
\end{equation}
where $D_S$ is the diffusion coefficient for molecules close to the surface and $\rho$ the molecular  density of the liquid (number per unit volume).  For regular liquids such as water, the result, Eq.~\eqref{rucken},  leads to molecular size slip length and suggests that the only way to get larger slip lengths would be for liquid molecules close to the surface to slip over a gas gap \cite{ruckenstein83}.

\section{Perspective} 
\label{perspective}

Because of the great advances in micro- and nano- fabrication technologies, the ability to engineer slip could have dramatic influences on flow since the viscous dominated motion can lead to large pressure drops and large axial dispersion. As was shown in this chapter, the small-scale interactions between a liquid and a solid  leads to extremely rich possibilities for slip behavior, with dependence on factors such as wetting conditions, shear rate, pressure, surface charge, surface roughness and dissolved gas.
 
We conclude this chapter by presenting a perspective summarizing the interpretation and use of the slip boundary condition (Eq.~\ref{bc}) to describe the motion at a liquid-solid boundary. 

\begin{itemize}
\item Physically, there is a difference between three different types of slip: (a) Microscopic slip at the scale of individual molecules, (b) actual continuum slip at a liquid-solid boundary ({\it i.e.}, beyond a few molecular layers) and (c) apparent (and effective) slip due to the motion over complex and heterogeneous boundaries. 

\item From a practical standpoint however, the distinction is not important. Whether it is real slip or apparent slip due to the interplay of many physical parameters, we have seen in this chapter that a large number of (generally small) experimental systems display some form of reduced resistance to fluid motion.

\item The (apparent) slip lengths reported experimentally span many orders of magnitude, from molecular lengths up to hundreds of nanometers. 
The impact of slip on systems with typical dimensions larger than tens of microns will therefore likely be limited. 

\item Molecular theories are able to predict  intrinsic slip lengths of up to tens of nanometers for hydrophobic systems, suggesting that any measurement of larger slip is affected by factors other than
purely fluid dynamical.

\item The parameters that contribute to apparent slip include roughness-induced dewetting, the amount and nature of dissolved gas, contamination by impurities and viscous heating. Other parameters that influence the magnitude of apparent slip include  contact angle, shear rate, electrical properties and pressure.

\item Finally, although it is usually assumed that slip only occurs on hydrophobic surfaces, a large variety of hydrophilic surfaces with different wetting properties have be shown to be prone to slip (Fig.~\ref{slip_depends}).

\end{itemize}

Other more complex behaviors remain to be understood, including dependence of the results on the molecular shape and size \cite{cheng02,galea04,leger03,granick02_langmuir}, probe size \cite{lumma03}, or  viscosity  \cite{craig01,neto03}. The development of alternative direct experimental methods would allow for a more precise quantification of slip phenomena. Similarly, it might be valuable to reproduce some of the experiments discussed above in degassed and clean environments to quantify the influence of dissolved gas on apparent slip. Answers to these questions will probably allow for a precise engineering of slip in small-scale systems.

\section*{Acknowledgments}

We thank L. Bocquet, K. Breuer, E. Charlaix, H. Chen, C. Cottin-Bizonne, B. Cross, R. Horn, J. Israelachvili, J. Klein,  J. Koplik,  J. Rothstein, T. Squires,  A. Steinberger, O. Vinogradova, and A. Yarin for useful feedback on an early draft of this chapter. Funding by the NSF Division of Mathematical Sciences, the Office of Naval Research and the Harvard MRSEC is acknowledged.

\bibliographystyle{plain}
\bibliography{bib_thesis_elauga}
\end{document}